 \documentclass[prl,aps,preprint,superscriptaddress]{revtex4-1}
\usepackage{amsmath, amssymb}
\usepackage[sort&compress]{natbib}
\usepackage{graphicx}
\usepackage{dcolumn}
\usepackage{bm}
\usepackage{hyperref}
\begin{document}
\title{Direct Observation of Interface and Nanoscale Compositional Modulation in Ternary III-As Heterostructure Nanowires}

\author{Sriram Venkatesan}
\email{sriram.venkatesan@cup.uni-muenchen.de}
\affiliation{Department of Chemistry and Center for NanoScience, Ludwig-Maximilians-Universit\"{a}t  M$\ddot{\mathrm{u}}$nchen, Butenandstr 5-13(E), 81377, M$\ddot{\mathrm{u}}$nchen,Germany} 
\author{Morten  H. Madsen}
\affiliation{Nano-Science Center and Center for Quantum Devices, Niels Bohr Institute, University of Copenhagen, 2100, Copenhagen, Denmark} 
\author{Herbert Schmid}
\affiliation {INM-Leibniz Institute for New Materials, 66123, Saarbr$\ddot{\mathrm{u}}$cken, Germany}
\author{Peter Krogstrup}
\author{Erik Johnson}
\affiliation{Nano-Science Center and Center for Quantum Devices, Niels Bohr Institute, University of Copenhagen, 2100, Copenhagen, Denmark}
\author{Christina Scheu}
\affiliation{Department of Chemistry and Center for NanoScience, Ludwig-Maximilians-Universit\"{a}t  M$\ddot{\mathrm{u}}$nchen, Butenandstr 5-13(E), 81377, M$\ddot{\mathrm{u}}$nchen,Germany}
%%%%%%%%%%%%%
\begin{abstract}
 Straight, axial InAs nanowire with multiple segments of Ga$_{x}$In$_{1-x}$As were grown. High resolution X-ray energy-dispersive spectroscopy (EDS) mapping reveal the distribution of group III atoms at the axial interfaces and at the sidewalls. Significant Ga enrichment, accompanied by a structural change is observed at the Ga$_{x}$In$_{1-x}$As/InAs interfaces and a higher Ga concentration for the early grown Ga$_{x}$In$_{1-x}$As segments. The elemental map and EDS line profile infer Ga enrichment at the facet junctions between the sidewalls. The relative chemical potentials of ternary alloys and the thermodynamic driving force for liquid to solid transition explains the growth mechanisms behind the enrichment.
\end{abstract}
%%%
\maketitle
%%%%%%%
Quasi-one-dimensional semiconducting nanostructures attract enormous attention owing to their physical properties ~\cite{Persson2004} and their potential application in future electronic devices.~\cite{Gudiksen2002,Wu2002,Huang2001,Hoffman2009} In particular, heterostructures ~\cite{Bjork2002,Borgstrom2006} draw considerable research interest because of the possibility to tune their band gap by compositional changes or by the ordered boundary defects like twin plane superlattices.~\cite{Ikonic1993,Ikonic1994,Caroff2009} Axial superlattice nanowires have been produced by interchanging group V elements (i.e.As and P).~\cite{Bjork2002,Maria2007,Jabeen2011} However, achieving axial superlattice growth of III-V elements by interchanging group III elements is difficult for some material combinations due to dissimilar interface energies.~\cite{Paladugu2008,Paladugu2007,Wallentin2010,Kimberley2007} Heterostructured nanowires intended for devices, the interfaces are of crucial importance.~\cite{Tomioka2012} Ternary heterostructures with phosphorus and antimony containing III-V compounds and binary III-V compounds such as GaAs on top of InAs have been reported frequently to be abrupt.~\cite{Maria2007,Chris1987,Dheeraj2008,Sven2005,Ohlsson2002} Whereas the opposite interface, i.e. InAs on GaAs are diffuse and/or kinked.~\cite{Paladugu2007,Kimberley2007} Only in a recent work by Messing et al. it has been shown that conditions can be controlled for axial growth of InAs on top of GaAs to obtain sharp interface.~\cite{Messing2011} Also, the growth of high quality,straight InAs/GaAs~\cite{Kimberley2012} and InAs/Ga$_{x}$In$_{1-x}$As/InAs heterostructure nanowire have been reported.~\cite{Peter2009} However,the inherent kinetics of the growth processes are not yet understood. In the present study, we report on the interfacial structure and composition of Au catalyzed InAs nanowires with multiple Ga$_{x}$In$_{1-x}$As segments grown periodically over a distance of more than 5 $\mu$m. These multi-segment heterostructures were investigated by advanced analytical transmission electron microscopy (TEM) /Scanning TEM (STEM) methods. Elemental mapping of the ternary barriers combined with thermodynamical calculations of the chemical potential of the compounds in the liquid Au catalyst particles provide insight into the growth mechanism. 

\par Nanowires were grown on epi-ready InAs (111) B wafers (Semiconductor Wafer, Inc.) in a Varian Gen-II molecular beam epitaxy system. Detailed discussion on growth parameters and procedures were explained elsewhere.~\cite{Madsen2012} Conventional TEM experiments were performed in a FEI Titan 80-300 S/TEM operated at 300 kV. Atomic resolution high-angle annular dark-field (HAADF) Z-contrast imaging and high resolution X-ray energy-dispersive spectroscopy (EDS)/spectroscopic imaging by X-rays (SIX) experiments were carried out using an advanced probe Cs-corrected analytical TEM/STEM system JEM-ARM 200CF with cold field-emission gun (C-FEG) operated at 200 kV.

\par Mean dimensions of the nanowires are 4.3 $\pm$1.0 $\mu$m in length, and 57 $\pm$9 nm in diameter, ($\pm$ indicate the standard deviation for 72 nanowires) were obtained from scanning electron microscopy images (not shown). A density of about 2 nanowires/$\mu$m${^2}$ was observed. The HAADF-STEM image in Fig 1a show a nanowire with a periodic superlattice of Ga$_{x}$In$_{1-x}$As segments in an InAs nanowire. Ga$_{x}$In$_{1-x}$As with a lower average atomic number in comparison with InAs give rise to dark contrast regions. The base of the wire has a diameter of $\sim$65 nm with a small decrease in the apex region to $\sim$50 nm giving a slight tapering. The length of the Ga$_{x}$In$_{1-x}$As segments is fairly constant throughout the wire whereas the length of the InAs regions increases from the bottom of the wire to the tip close to the Au catalyst.~\cite{Peter2009} The majority of the wires grow straight ($\sim$85 \%) without kinking. Maintaining a Ga concentration corresponding to x$\sim$0.35 in each Ga$_{x}$In$_{1-x}$As segment ensures a high fraction of straight axially grown nanowires.~\cite{Peter2009,Madsen2012} The wires predominantly exhibits the hexagonal wurtzite (WZ) structure and the preferential growth direction is along the hexagonal closed packed [0001] WZ as inferred from electron diffraction data shown in figure 2b.~\cite{Kimberley2012,Hadas2009,Galicka2008,Krogstrup2012} 

\par Our analysis of the multiple Ga$_{x}$In$_{1-x}$As segments in the heterostructures revealed the following picture. At first sight, an enhanced dark contrast region is visible at the Ga$_{x}$In$_{1-x}$As/InAs interface in Z-contrast HAADF-STEM imaging as shown in figure 1b. The contrast change at this interface is seen in all the segments of the wires. Atomic resolution HAADF-STEM image (Fig 1c) at the interface revealed a clear presence of about 10-15 atomic layers of ZB structure with preceding and succeeding WZ structure. However, no pronounced contrast difference or Ga enrichment was noticed at the InAs/Ga$_{x}$In$_{1-x}$As interface implying a relatively sharp interface. Under ideal HAADF imaging conditions, difference in the contrast can be directly correlated to a notable change in composition. Post growth EDS/SIX elemental mapping was performed to estimate the element distribution and to understand the growth behavior. Elemental line profiles along the segment (integrated over 60 pixels as indicated in Fig.2a) are shown in Fig.2d which provides a qualitative correlation between Z-contrast and composition. The Ga map (Fig.2e and the overlay image Fig.2h) reveal a significant Ga enrichment at the Ga$_{x}$In$_{1-x}$As/InAs dark interface region. The maximum of Ga signal in the line profile (indicated by red arrow in Fig 2d) coincides with the dark contrast region in HAADF image of figure 1b. The Ga enrichment occurs at expense of In, visualized as deficiency in the In map (Fig.2f) and as a dip at the same region in the line profile (Fig. 2d). As expected, the As content remains constant along the whole wire.

\par For a better understanding of the growth, the top three and bottom three Ga$_{x}$In$_{1-x}$As segments from five nanowires were analyzed. If changes occur, they are most pronounced in these two regions, that represent segments formed at the beginning and end of the growth (near the substrate and Au catalyst). Selected regions of interest (InAs, InAs/Ga$_{x}$In$_{1-x}$As interface region, Ga$_{x}$In$_{1-x}$As segment and Ga$_{x}$In$_{1-x}$As/InAs interface region) were quantitatively evaluated with EDS. Measurements were performed locally on the dark Ga enrichment region for several Ga$_{x}$In$_{1-x}$As/InAs interfaces of different wires to get an average Ga composition value. The Ga content in the bottom region Ga$_{x}$In$_{1-x}$As segments corresponds to x= 0.45 $\pm$0.06 while the top region segments yields x= 0.20 $\pm$0.03. Similarly, the Ga content in the enriched interface at the bottom region segments corresponds to x=0.54 $\pm$0.04 and at the top region segments to x=0.34 $\pm$0.07. The In has been traded off for the Ga enrichment and the As content was constant at 50 atomic \% throughout the nanowires. Overall the Ga content is always higher for the bottom region segments than the segments at the top. 
 
\par The explanation for such a Ga enrichment has been often attributed to the stronger affinity of In to Au than Ga.~\cite{Peter2009,Messing2011} This argument is valid upon assuming that Ga alloys with the In-Au alloy catalyst and that the vapor-liquid-solid (VLS) growth happens via volume diffusion in the catalyst droplet. One reason for occurrence of wide transition regions is the introduction of a new element decreases the solubility limit of the existing element in the catalyst.~\cite{Kimberley2012} Since solubility is an equilibrium quantity, it is a reference measure of given element concentration in a given alloy. Therefore, we will discuss our results in terms of the thermodynamic driving force (difference in chemical potentials) for the liquid to solid transition which is a measure of how far from equilibrium a given component is during growth. Even though the liquid to solid transition rates can be thought of as being driven by thermodynamic driving forces toward equilibrium, it is not possible to directly relate such driving forces with steady state concentrations of the involved elements during growth. This is because of the high pressure supply from the adatoms and gas phases. Additionally, the transition rates are not only determined by the thermodynamic driving forces but also by prefactors that do not directly involve thermodynamic parameters.~\cite{Peter2013} However, we can use driving forces as a measure of how far from equilibrium the liquid to solid transition for each group III element is, and use it to explain the observed compositional distribution. If we introduce an element which quickly increases the driving force, a smaller amount of the material is absorbed before it reaches its steady state composition and therefore will imply a sharp transition region.~\cite{Li2008} This seems to be the case for the InAs/Ga$_{x}$In$_{1-x}$As interface region where Ga is introduced. In contrast, the Ga$_{x}$In$_{1-x}$As/InAs interface is much wider because the In atoms seem to have a smaller driving force towards the solid than Ga, and Ga is expelled from the liquid by solidification into the nanowire, prior to reaching the steady state liquid composition.
 
\par We validate these arguments by analyzing the thermodynamics of the quaternary liquid. As an indication of the affinities of In and Ga in the liquid, we first use chemical potentials of the ternary Au-In-As and Au-Ga-As alloys, and consider the liquid to solid thermodynamic driving forces required to form pairs of solid Ga-As and In-As from the liquid, using thermodynamic data derived by Glas.~\cite{Glas2010} A high driving force means that the affinity of the relevant species is low in the liquid phase (or high for the liquid to solid reaction affinity). The driving forces have been plotted for a given In and Ga mole fraction of 0.35 as a function of As mole fraction in the liquid droplet at the growth temperature of T = 425 $^{\circ}$C (see figure 3a). The plot shows that a GaAs pair in the liquid has a higher thermodynamic driving force towards the solid phase than an InAs pair. For the calculation it is assumed that the Au remains in the droplet and that the stoichiometry of the solid is fixed to 1:1 (In/Ga:As), which are reasonable assumptions. To determine the relation between the two group III elements in the liquid, we show in figure 3b a hypothetical scheme of how the system may go towards minimum driving forces under the given constrains. The red circle in the figure illustrates a region corresponding to the applied group III flux ratio which could roughly be equal to the group III compositions in the liquid during growth. A measure of the liquid chemical potentials during growth are here labeled $\mu^G_{l,III-V}$. When the shutters are closed the Ga will be expelled, and the system moves towards minimum driving force conditions with liquid chemical potentials $\mu^M_{l,III-V}$. Thus, once the In and Ga shutters are closed simultaneously, the system moves towards thermodynamic equilibrium by expelling primarily Ga-As pairs from the liquid. This explains the high concentration of Ga at the termination of the Ga$_{x}$In$_{1-x}$As segment. As we have shown, the thermodynamic driving forces explain the reasons behind the Ga enrichment at the Ga$_{x}$In$_{1-x}$As/InAs interface, which subsequently results in the formation of ZB region.

\par The investigation on the sidewalls of the wire show clear Ga presence (see figure 2h) with a In depletion and constant As. Upon careful analysis, the EDS Ga line profiles across the wire (see figure 2c) reveals that the Ga signal (red) is enhanced at four regions in the wire, which can also be seen in the Ga elemental map shown in figure 2e. A schematic representing this scenario i.e a thin GaAs shell with Ga enrichment at the facet edges is shown in figure 4. The Ga intensity profile from the scheme is seen to match the experimental observed profile (Figure 2c). Similar results have been reported for core shell nanowires where enrichment is observed at the side wall facet junctions.~\cite{Rudolph2013,Wagner2010} The Ga enrichment at the junctions connecting the side walls could be attributed to the related change in the chemical potential gradient, the growth and the entropy of mixing effects.~\cite{Biasiol1998} More investigations are required to understand this process in detail. Another observation is that, about 10 atomic \% Ga is incorporated within InAs section of the bottom region, while no Ga has been detected in the top InAs barrier, though the Ga shutter was opened consistently for 10 seconds for growing each of the segments. A lower Ga sidewall diffusion length of 0.25 $\mu$m for later grown segments compared to 0.8 $\mu$m for early grown segments was calculated in our recent work and it supports our present observations.~\cite{Madsen2012} In general, In-Ga-As-Au alloy is not an ideal solution system; there is a strong influence of one to each other implies elemental interactions on their solubilities depending on the ratio between the elements and time of introduction. These factors play a critical role in determining the conditions in the alloy and thus the quality of the interface. Although the VLS growth mechanism is commonly accepted, depending on the temperature, species involved and other growth parameters the mechanism will undergo slight modification. We assume that VLS growth and interface/side wall diffusion are the major incorporation mechanisms in our growth.
 
\par In summary, we have demonstrated the growth of multiple Ga$_{x}$In$_{1-x}$As segments into an InAs nanowire and analyzed them in detail. Our systematic study on compositions reveals a clear difference in Ga concentration between the early grown Ga$_{x}$In$_{1-x}$As segments close to the substrate and the latest grown Ga$_{x}$In$_{1-x}$As segments near the Au catalyst. A transition region with an increased Ga content accompained by a zinc blende structure of about 4 nm is observed at all the Ga$_{x}$In$_{1-x}$As/InAs interfaces. Also a GaAs shell and Ga enrichment was observed at the side facet edges of the hexagonal nanowire. The occurrence of Ga enrichment is explained using thermodynamical concepts. We have shown that the growth conditions for multiple segments of Ga$_{x}$In$_{1-x}$As in InAs nanowires change significantly during growth and differs from growth of single or double segments reported earlier.~\cite{Peter2009}  In this framework, our work may provide a better understanding to tailor periodicities and stoichiometry of III-V nanowire heterostructure containing multiple electronic barriers involving In and Ga. Finally, this work opens up the possibility to tune the band gap of individual segments of Ga$_{x}$In$_{1-x}$As by varying the Ga concentration within a nanowire.

\par S.V and C.S would like to thank the German research foundation (DFG) for funding in the Cluster of Excellence "Nanosystems Initiative Munich" (NIM) and Dr. Markus {D$\mathrm{\ddot{o}}$blinger} for providing assistance with the microscope. M.H.M and P.K acknowledge the financial support from the Danish Strategic Research Council, the Advanced Technology Foundation through Project 002-2009-1, and EU FP7 project SE2ND.

%%%%%%%%%%%%%
%

\clearpage

\begin{figure}
  \begin{center}
	\includegraphics[width=1.00\textwidth]{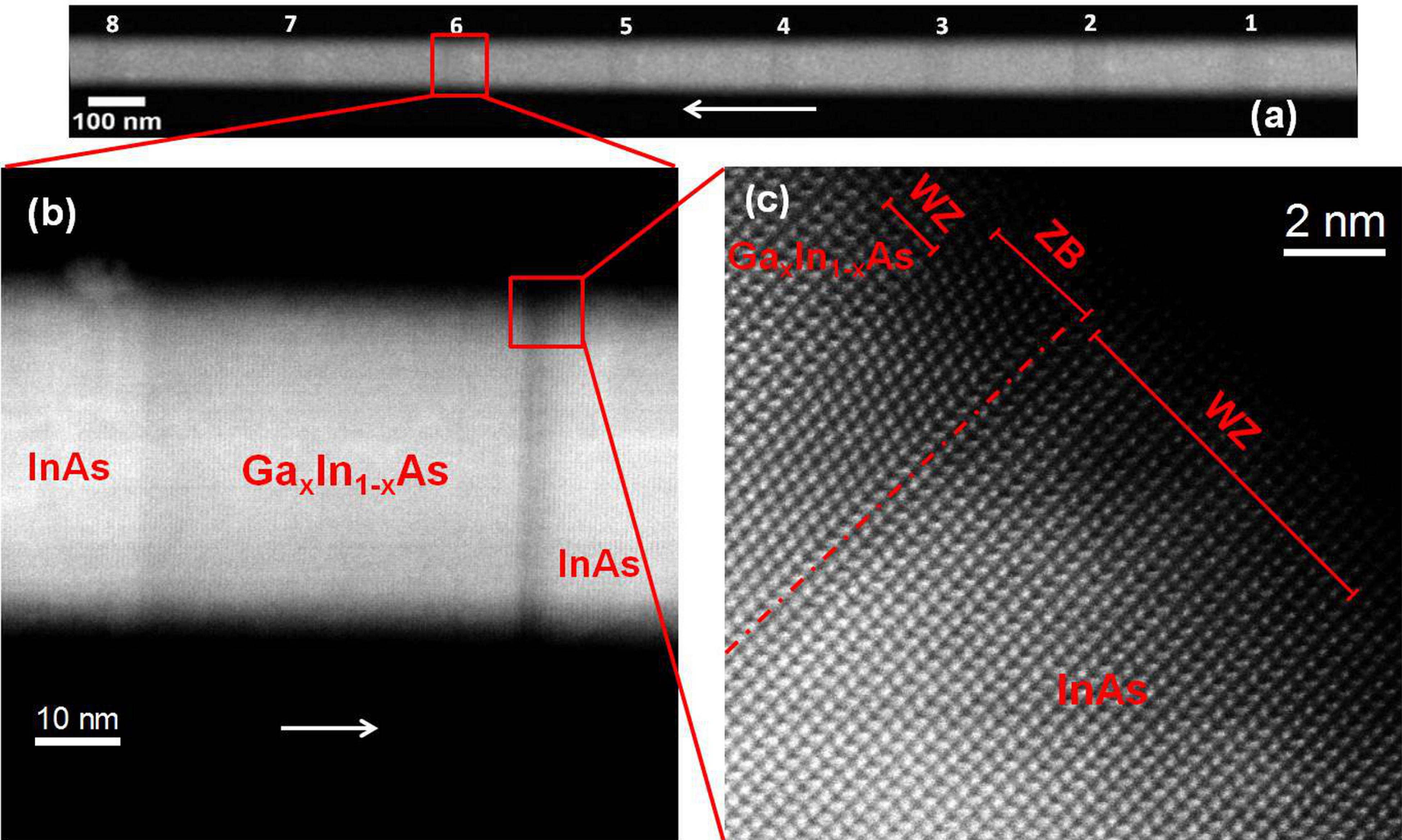}
\caption{(a) Overview HAADF-STEM image of the wire shows 8 of 11 Ga$_{x}$In$_{1-x}$As segments in an InAs nanowire.(b) Magnified view revealing enhanced dark contrast at the Ga$_{x}$In$_{1-x}$As/InAs interface. (c) Atomic resolution HAADF Z-contrast image showing the dark contrast Ga region with a ZB structure. Growth direction is indicated by white arrows.}
    \label{fig:Figure 1}
  \end{center}
\end{figure}

\vspace*{5mm}

\begin{figure}
  \begin{center}
	\includegraphics[width=0.8\textwidth]{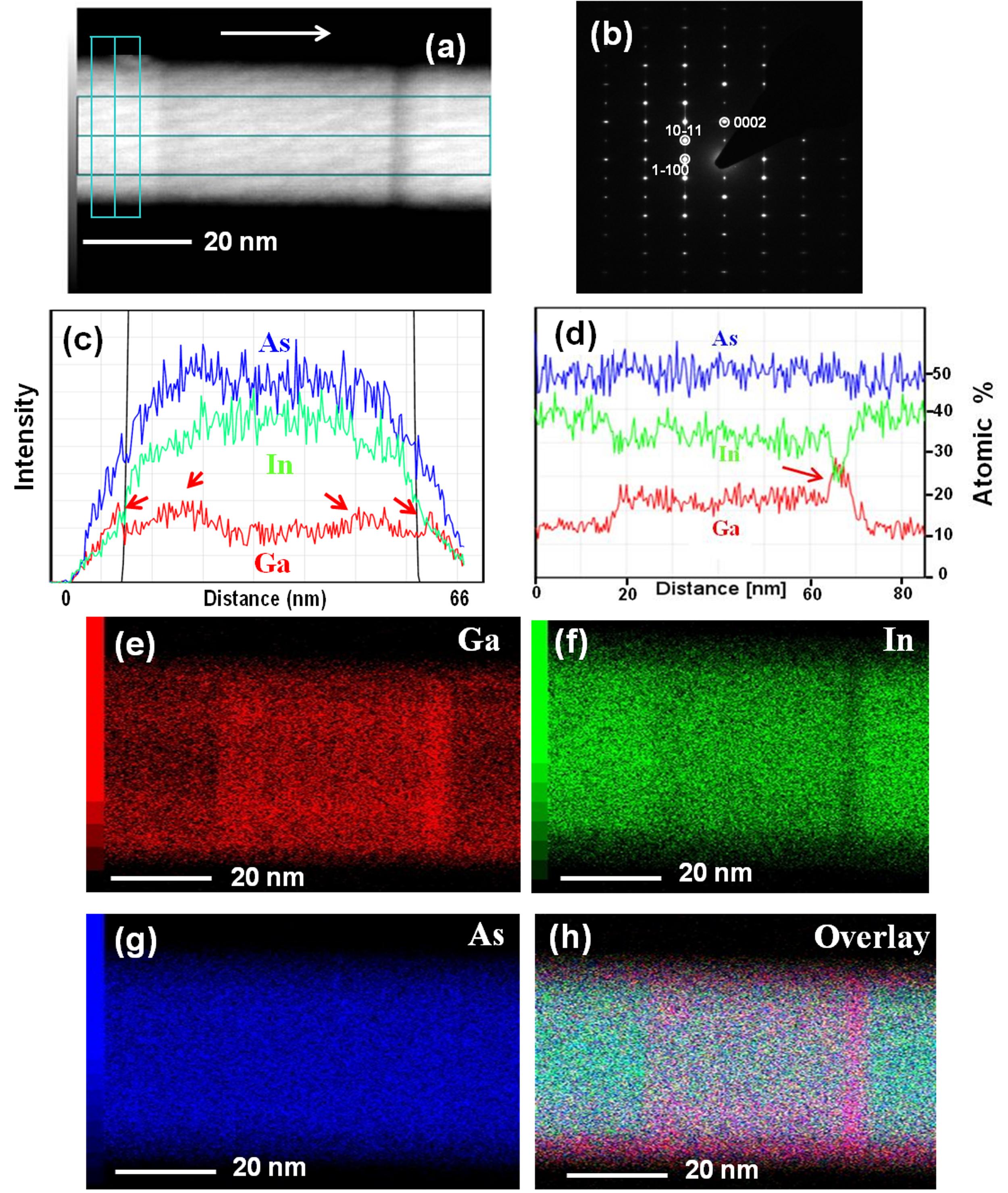}
\caption{(a) Electron diffraction pattern from a another wire with growth direction [0002] and electron beam direction [2-1-10] shows a WZ structure (b) HAADF-STEM image showing a Ga$_{x}$In$_{1-x}$As segment (number 11) in InAs nanowire where EDS mapping was performed (growth direction indicated by white arrow). (c) Line profiles of Ga, In and As obtained across the wire diameter and (d) along the growth direction revealing the enrichment of Ga (indicated by red arrows). Figure (e) (f) and (g) show EDS/SIX elemental maps of Ga, In and As, respectively. (h) Overlay image of Ga, In and As, showing the Ga enrichment at Ga$_{x}$In$_{1-x}$As/InAs interface and at the side wall. Note: The thin GaAs shell is not visible in (a) due to lower average atomic number and low thickness.}
    \label{fig:Figure 2}
  \end{center}
\end{figure}

\begin{figure}
  \begin{center}
	\includegraphics[width=0.8\textwidth]{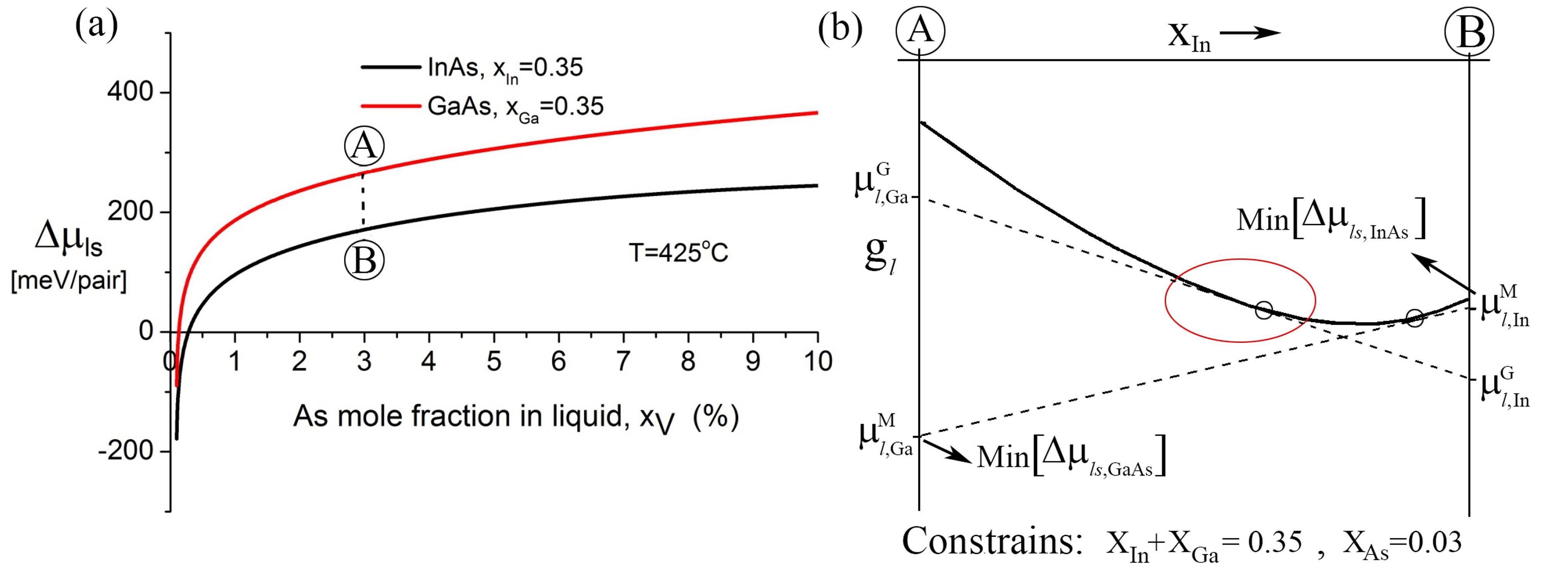}
\caption{(a) Relative chemical potentials of liquid and  with mole fractions of In and Ga in the Au droplet plotted against As mole fraction in the liquid for a growth temperature of 425 $^{\circ}$C. (b) A sketch showing how the ls driving forces change when varying the group III element. The solid line is the free energy curve and the tangents are used to find the chemical potentials on the respective axes.}
    \label{fig:Figure 3}
  \end{center}

\end{figure}
\begin{figure}
  \begin{center}
	\includegraphics[width=0.5\textwidth]{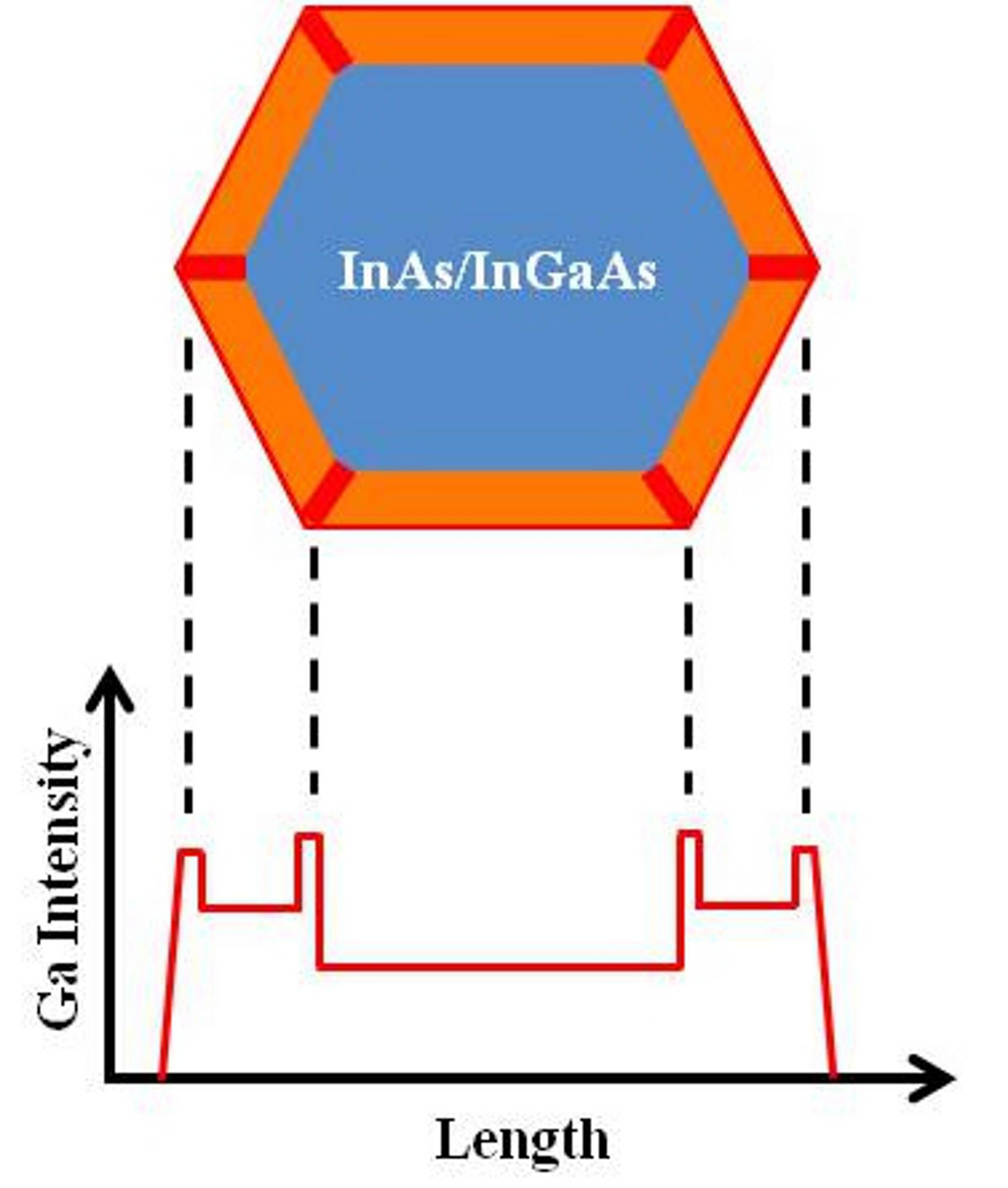}
\caption{(a) A schematic represents a thin GaAs shell is formed with a higher Ga concentration at the facets edges for the line profile in figure 2c. Note: The shape and size of the Ga enrichment junctions shown are for representation purpose and are not scalable.}
    \label{fig:Figure 4}
  \end{center}

\end{figure}
\end{document}